\documentclass[twocolumn,prb]{revtex4-2}
\usepackage{titlesec}
\usepackage{subfigure}
\usepackage{amsmath}
\usepackage{amsfonts}
\usepackage{amssymb}
\usepackage{feynmp}
\usepackage{graphicx}
\usepackage{setspace}
\usepackage{comment}
\usepackage{dsfont}
\usepackage{color}
\usepackage[pdftex,colorlinks=true,linkcolor=blue,citecolor=blue]{hyperref}

\begin{document}

\title{Topological metals constructed by sliding quantum wire arrays}

\author{Zheng-Wei Zuo}
\author{Linxi Lv}
\author{Dawei Kang}

\affiliation{School of Physics and Engineering, Henan University of Science and Technology, Luoyang 471023, China}
\date{\today}

\begin{abstract}
A general strategy of alternated slide construction to craft topological metals is proposed, where there is a relative slide between the odd and even chains in the trivial spinless quantum wire array. Firstly, taking the three-leg ladder as an example, we find that alternated slide can induce a topological phase transition from the normal metal to topological metal phases, which are protected by inversion symmetry. Remarkably, topological metal without nontrivial edge states is found, and the bulk-boundary correspondence breaks down. Secondly, the two-dimensional quantum wire arrays with alternated slide manifests similar physical behaviors. Two types of topological metal phases emerge, where there are gapless bulk bands with and without nontrivial edge states. These results could be confirmed by current experimental techniques.
\end{abstract}

\maketitle

\section{Introduction}
Nowadays, identifying, classifying, and engineering topological quantum matters is of extensive interest in condensed matter physics\cite{Hasan10RMP, QiXL11RMP, Alicea12RPP, Franz15RMP, ChiuCK16RMP, ArmitageNP18RMP}. The topological insulator, topological superconductors, and topological metal (semimetals) are prototype examples of topological phases. For the electronic materials, based on symmetry-indicator theory\cite{PoHC17NTC, WatanabeH18SA} and (magnetic) topological quantum chemistry\cite{BradlynB17NT,ElcoroL21NTC}, a myriad of topological materials\cite{ZhangTT19NT, VergnioryMG19NT, TangF19NT, XuYF20NT, Bernevig22NT} have been predicted and discovered. For the topological metals, there are many types depending on the dimensionality, dispersion (slope and order), and degeneracy of band crossing. For example, according to the dimensionality of nodal gap closing, topological metals can be divided into nodal-point, nodal-line, and nodal-surface topological metals. Based on the slope of band dispersion near the crossing, there are types I and II topological metals. Additionally, there is another topological metal, with the topological localized states with eigenvalues embedded in the continuum of extended states, dubbed topological bound states in the continuum\cite{YangBJ13NTC, XiaoYX17PRL, NiX19NTM, ChenZG19PRB, YingXZ18PRL, YingXZ19PRB, BahariM19PRB, BenalcazarW20PRB, CerjanA20PRL, LiZZ20PRL, JangjanM20SR, JangjanM21SR, JangjanM22PRB, LiuLH23PRL}. A variety of gapless symmetry-protected topological phases\cite{ScaffidiT17PRX, ParkerDE18PRB, JiangHC18SB, GoncalvesM19PRL, ThorngrenR21PRB, VerresenR21PRX, HidakaY22PRB} beyond the topological metals have garnered much attention recently.

There are many theoretical and experimental tools for engineering topological quantum matters. One of the flagships is that the magnetic field induces the quantum Hall effect in two-dimensional (2D) electron gas\cite{Klitzing80PRL,Hofstadter76PRB}. Combined the electron-electron interaction, the magnetic field can also cause the fractional quantum Hall effect\cite{Tsui82PRL, Laughlin83PRL}. The modulated hoppings and potentials provide another approach to realize different types of topological band phases. The Su-Schrieffer-Heeger (SSH) model\cite{SuWP79PRL}, Aubry-André-Harper (AAH) model\cite{LangLJ12PRL, GaneshanS13PRL}, and Kekulé distortion of graphene-like structures\cite{HouCY07PRL} are representative examples. On the other hand, the interplay of magnetic field and modulated hoppings could result in the high-order topological insulators\cite{Benalcazar17SCI, Benalcazar17PRB, ZuoZW21JPD}. More recently, by adjusting the twist angle, graphene moiré superlattices have been widely explored as platforms for exploring correlated physics, energy band topology, and superconductivity\cite{CaoY18NT1, CaoY18NT2}.

In this paper, we propose a simple scheme to construct topological metals by way of trivial spinless quantum wire arrays sliding against each other. Firstly, for the simple three-leg ladder, the center leg has a relative translation compared with the bottom and top chains. The ladder showcases unconventional properties. By way of a unitary transformation, the Hilbert space of the system could be separated into two independent subspaces, where a trivial subspace and a topological nontrivial subspace coexist. Because of the different nearest-neighbor interchain couplings (depending on interchain distance), the system undergoes a topological phase transition from normal metal to inversion symmetry-protected topological metal phases, where for the topological metal, the nontrivial edge states appear or not depending on the strength of intrachain couplings. Second, for the 2D quantum wire array with alternated slide, similar physical phases also emerge.

\begin{figure}[tbh]
\centering
\includegraphics[width=\columnwidth]{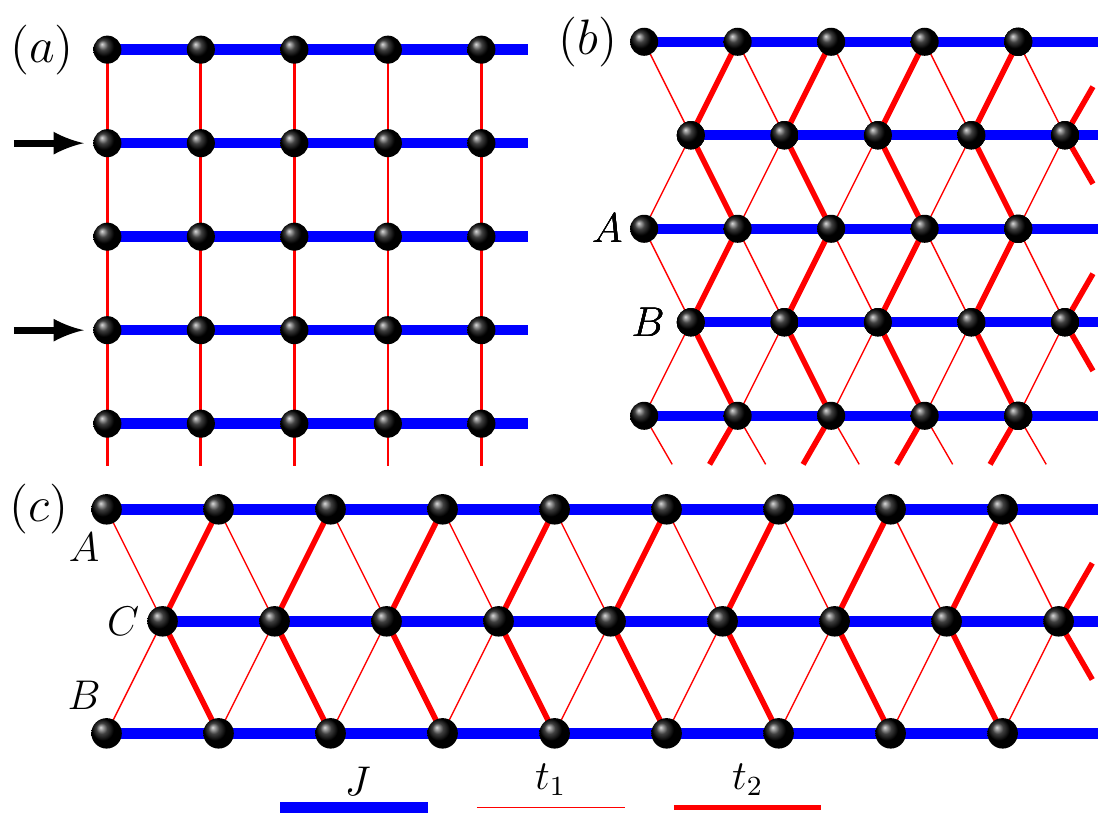}
\caption{(a) Schematic picture of 2D lattice system (trivial quantum wires array). (b) The modified 2D lattice system with quantum wires array sliding against each other. (c) The three-leg (labeled $A, B,$ and $C$) ladder. The hopping in quantum wires is $J$; the coupling strengths between quantum wires are $t_1$ and $t_2$.}
\label{Fig1}
\end{figure}

\section{Model}
Now one can consider a 2D lattice system of trivial spinless quantum wire arrays (Fig.\ref{Fig1}a). Then, the even quantum wires have a gradual radial-direction translation relative to the odd quantum wires, as shown in Fig.\ref{Fig1}b. For the nearest-neighbor interchain coupling, there are two different strength hoppings $t_1$ and $t_2$ (red thin and thick lines) depending on distance. As the translation of even quantum wires pushes ahead, the coupling strength $t_1$ ($t_2$) decreases (increases). When $t_1<t_2$, one may intuitively think there is nothing happening. However, a topological phase transition from normal metal to topological metal takes place, which is an unexpected and counterintuitive theoretical discovery. We unveil two types of topological metals, where the nontrivial edge states can appear or not. For the strong intrachain coupling $J$, the system could enter into the topological metal phase without nontrivial edge states. When the even quantum wires are shifted from one lattice constant, the system returns to the 2D normal quantum wire arrays. Furthermore, these results could be confirmed experimentally in a photonic waveguide array, photonic crystals, topoelectrical circuits, or coupled acoustic resonators.  For example, the coupled-waveguide arrays are manufactured by the femtosecond laser direct-writing technique. The coupling coefficients $J$, $t_1$, and $t_2$ are determined by the evanescent mode coupling between adjacent waveguides, which can be tailored by waveguide spacing.

\subsection{Three-leg ladder}
For simplicity, we use the three-leg ladder (Fig.\ref{Fig1}c), which serves as a starting point. We can write the tight-binding model (set lattice constant $a=1$) as
\begin{equation}
H=\sum_{j, \alpha, \beta}^L J c_{\alpha , j}^{\dagger}c_{\alpha , j+1}+t_1 c_{\beta , j}^{\dagger}c_{C , j}+t_2 c_{\beta , j+1}^{\dagger}c_{C, j}+H.c.,
\end{equation}
where $c_{\alpha , j}^{\dagger}$ ($c_{\alpha , j}$) is the fermionic creation (annihilation) operator at lattice site $j$ in the leg $\alpha$. Here $\alpha$ stands for the legs $A, B,$ or $C$, and $\beta$ denotes the legs $A,$ or $B$. The $J$, and $t_1$ ($t_2$) are strengths of the intrachain coupling and interchain coupling, respectively. The length of the chain is $L$.
 
Applying the Fourier transform, we can express the bulk momentum-space Hamiltonian as
(base ($c_{A}^{\dagger}$ , $c_{B}^{\dagger}$, $c_{C}^{\dagger}$))
\begin{equation}
H_{o}=\left(
\begin{array}[c]{ccc}
2J\cos k_x & 0 & t_1+t_2 e^{ik_x}\\
0 & 2J\cos k_x & t_1+t_2 e^{ik_x}\\
t_1+t_2 e^{-ik_x} & t_1+t_2e^{-ik_x} & 2J\cos k_x
\end{array}
\right)
\end{equation}
If we use a unitary transformation $H_D=Q^{-1}H_{o}Q$, the Hamiltonian can be block-diagonalized
\begin{align}
H_D &  =\left(
\begin{array}[c]{ccc}
2J\cos k_x & 0 & 0\\
0 & 2J\cos k_x & \sqrt{2}\left(  t_1+t_2e^{ik_x}\right)  \\
0 & \sqrt{2}\left(  t_1+t_2e^{-ik_x}\right)   & 2J\cos k_x
\end{array}
\right)\label{Hqtotal}
\end{align}
where the unitary matrix
\begin{equation}
Q=\frac{1}{\sqrt{2}}\left(
\begin{array}[c]{ccc}
-1 & 1 & 0\\
1 & 1 & 0\\
0 & 0 & \sqrt{2}
\end{array}
\right)
\end{equation}

The Hamiltonian divides into two independent blocks $h_1=2J\cos k_x$ and $h_2=\left(
\begin{array}[c]{cc}
2J\cos k_x & \sqrt{2}\left(  t_1+t_2e^{ik_x}\right)  \\
\sqrt{2}\left(t_1+t_2e^{-ik_x}\right)   & 2J\cos k_x
\end{array}
\right)$.  Thus, the Hilbert space becomes separable. We can diagonalize the Hamiltonian in Eq.\ref{Hqtotal} and obtain the energy eigenvalues. The energy dispersions are
\begin{align}
E_1 &  =2J\cos k_x,\\
E_{\pm} &  =2J\cos k_x\pm\sqrt{2\left(  t_1^2+t_2^2+2t_1t_2\cos k_x\right) } \label{Epm}
\end{align}

The lower $2\times2$ block $h_2$ is equivalent to the SSH model with next-nearest-neighbor (NNN) hopping\cite{LonghiS18OL,BeatrizP19PRB, JiaoZQ21PRL}, where the chiral symmetry is broken and the inversion symmetry is conserved. It means that the block-diagonalization separates the Hilbert space of the system into two independently subspaces, where a trivial subspace and a topological nontrivial subspace coexist. The system can be viewed as a hybrid of a trivial quantum wire and a virtual SSH model with NNN hopping. Thus, the whole system could enter into the topological metal phase, where the gapless bulk states with topological bound edge states appear. Here, let us firstly investigate the topological properties of the subspace $h_2$. The system displays the inversion symmetry $\mathcal{I}$ where $\mathcal{I}h_2(k_x)\mathcal{I}^{-1}=h_2(-k_x)$, where the inversion symmetry operator $\mathcal{I}$ is defined by
\begin{equation}
\mathcal{I}=\left(
\begin{array}[c]{cc}
0 & 1 \\
1 & 0
\end{array}
\right)
\end{equation}
and $h_{2}(k_x)$ commutes with $\mathcal{I}$ at inversion-invariant momenta $k_x=\left(0,\pi\right) $. Thus, eigenstates of $h_{2}(k_x)$ have a well-defined parity $\zeta_{i}\left(k_{\text{inv}}\right)=\pm1$ at those points. According to the parity value of inversion symmetry $\mathcal{I}$, we can calculate the topological invariant $\mathcal{N}$\cite{Hughes11PRB}(the Zak/Berry phase\cite{Zak89PRL} can also characterize this topological phase transition)
\begin{equation}
\mathcal{N}=\left\vert n_1-n_2\right\vert
\end{equation}
where $n_1$, $n_2$ are the number of negative parities at $k_x=0$ and $k_x=\pi$, respectively. So we get $\mathcal{N}=1$ (topological phase, the Zak phase is $\pi$) for $t_1<t_2$ and the two degenerate edge states would appear in a finite system. For $t_1>t_2$, the topological invariant $\mathcal{N}=0$ and the subspace $h_2$ is trivial.

It is worthwhile noting that the topological phase transition is independent of the intrachain hopping $J$. The intrachain hopping $J$ term in subspace $h_2$ is proportional to the identity matrix and does not change the bulk eigenstates. The eigenfunctions of $h_2$ can be explicitly expressed as $u_\pm=(1, \pm e^{-i\phi})^T$ ($\phi$ is the phase angle of complex number $t_1+t_2e^{ik_x}$), which is independent of intrachain hopping $J$. Thus, the topological invariants ($\mathcal{N}$, and Zak phase) are fixed for any values of intrachain hopping. However, for strong intrachain hopping $J$, the energy gap (see Hamiltonian $h_2$ and Eq.\ref{Epm}) could close without band crossing owing to the indirect nature of the gap (see Appendix \ref{AppendixA} for more details). At the same time, in a finite system, the two degenerate edge states would assimilate into the bulk bands and become delocalize. Thus, the conventional bulk-boundary correspondence breaks down. The subsystem is still in the topological phase (without the nontrivial edge states). For the entire system, the topological bands ($h_2$) coexist with the single trivial bulk band ($h_1$). Thus, there are two types of topological metals, where the nontrivial edge states can appear or not.

\begin{figure}[t]
\centering
\includegraphics[width=\columnwidth]{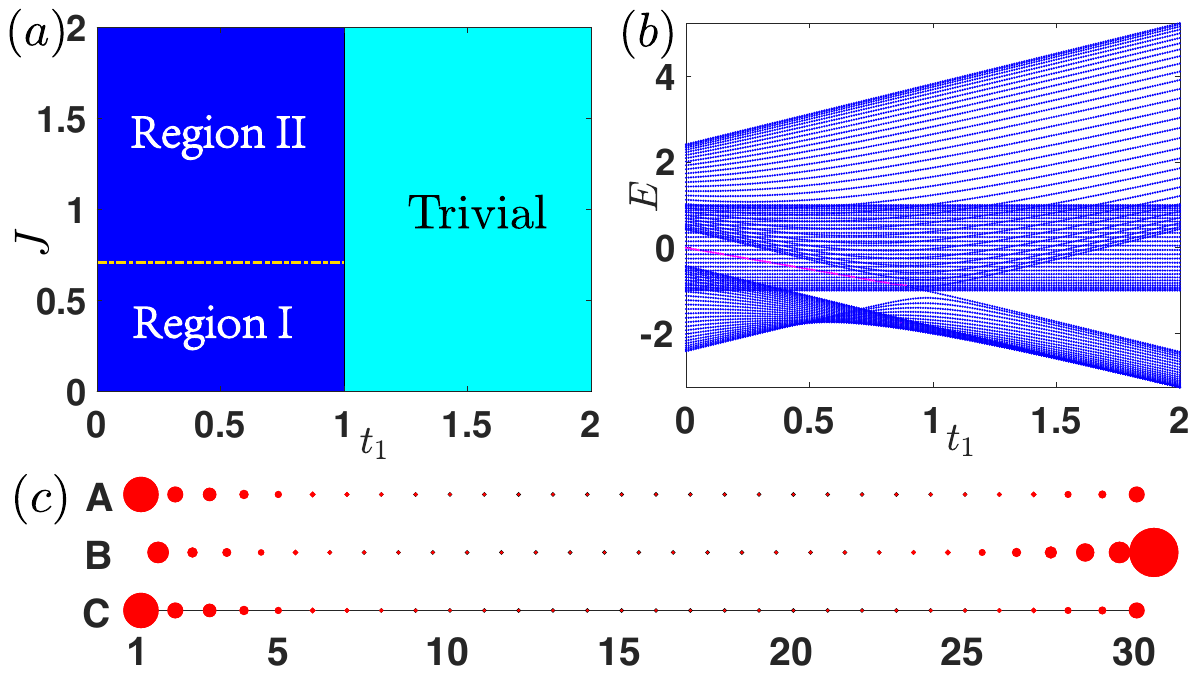}
\caption{(a) The phase diagram of the three-leg ladder with $t_2=1$. Region I (II) represents that the system is in the topological metal with (without) bound end states. Other regions are the trivial normal metal. (b) Energy spectra as a function of hopping $t_1$ with system size $L=30$, $t_2=1$, and $J=0.5$ under open boundary condition. The red dots indicate the topological bound end states identified by IPR. (c) The spatial distribution of the two topological bound end states; other parameters are $L=30$, $t_2=1$, and $t_1=J=0.5$.}
\label{Fig2}
\end{figure}

\subsection{Phase diagram}
According to the behaviors of the edge states and topological invariant $\mathcal{N}$ in $h_2$ subspace, we can numerically obtain the phase diagram of the whole system (set $t_2=1$ here) shown in Fig.\ref{Fig2}a. It shows that, as the interchain coupling $t_1$ decreases (the slide displacement increases), the system changes from the normal metal to topological metal. When the slide displacement is larger (smaller) than half of one lattice constant, the system is in topological (normal) metal states. For the interchain couplings $t_1<t_2$, the system evolves from the topological metal with nontrivial edge states to the topological metal without edge states (region I to region II) when the hopping $J$ increase (the critical value is $\sqrt2/2$ calculated by the edge state energy curve and the energy dispersions $E_\pm$). For the interchain couplings $t_1>t_2$, the system is in trivial normal metal phase. 

Next, let us investigate the topological phase transition. The energy spectra under open boundary condition are plotted in Fig.\ref{Fig2}b with regard to the hopping $t_1$ when system size $L=30$, $t_2=1$, and $J=0.5$.  To identify the topological edge (localized) states from bulk states, we use the inverse participation ratio (IPR), which is defined by $\mathrm{IPR_n}=\sum_{j=1}^{3L} \left|\psi_j(n)\right|^4$. The IPR of an extended state scales as $1/3L$, thereby vanishing in the thermodynamic limit, while remaining finite for a localized state. The red dots in Fig.\ref{Fig2}b show the topological bound end states identified by $\mathrm{IPR_n}$. Thus, the nontrivial edge states coexist with gapless bulk states at the Fermi level. At the same time, they do not couple with each other because they belong to different subspaces and the two subspaces are orthogonal. As the hopping $t_1$ increases, the system changes from the topological metal with nontrivial edge states to normal metal. As shown in Fig.\ref{Fig2}c, the two nontrivial edge states with $L=30$, $t_2=1$, and $t_1=J=0.5$ are clearly localized in the left and right boundaries. This type of topological metal is also identified by other systems\cite{XiaoYX17PRL, JangjanM20SR} (known for topological bound states in the continuum).
 
\begin{figure}[t]
\centering
\includegraphics[width=\columnwidth]{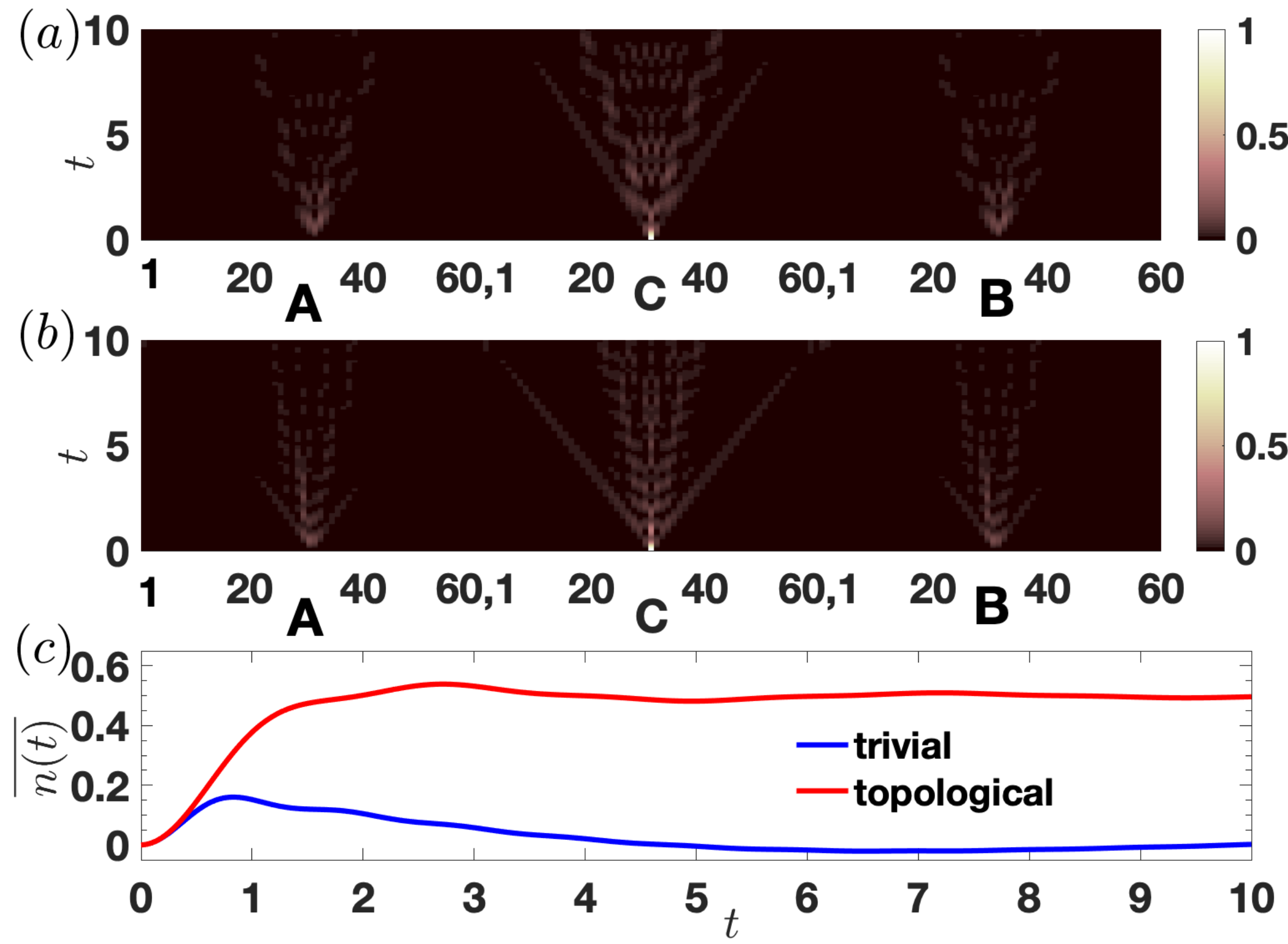}
\caption{Single-particle quantum walks on the three-leg (labeled $A$, $B$, and $C$) ladder with $J=1$, $t_2=1$, and $L=60$ under open boundary condition. The quantum walker is initially positioned on the center of leg $C$. (a) Topological metal phase case without the two bound end states with hopping $t_1=0.5$. (b) Normal metal case with hopping $t_1=1.2$. (c) Averaged mean density displacements for the topological metal (red line) and normal metal (blue line) cases.}
\label{Fig3}
\end{figure}
 
For the topological metals without nontrivial edge bound states, it is difficult to distinguish them from the normal metal because the bulk-boundary correspondence breaks down. Here, we use the experimentally continuous-time quantum walks\cite{VenegasAndraca12} to decode the topological properties of topological metal without the edge bound states.  Firstly, a single-particle state $\left|\Psi_0\right\rangle$ is initially located at the center of leg $C$.  The time-dependent density distribution of the single-particle quantum walker is given by $n(t)=\sum_n n(|A_n(t)|^2+|B_n(t)|^2+|C_n(t)|^2)$, where $A_n(t)$ ($B_n(t)$ and $C_n(t)$) is the occupation amplitudes of the dynamical evolution state $|\Psi(t)\rangle$ in the $n$-th unit cell of the leg A (B and C) sites, where $|\Psi(t)\rangle=e^{-iHt}\left|\Psi_0\right\rangle$. The time-averaged mean displacement in the time interval $(0, T)$ can be calculated by $\overline{n(t)}=\int_0^T n(t)dt/T$. For large $T$,  $\overline{n(t)}$ approaches a constant asymptotic value\cite{LonghiS18OL, JiaoZQ21PRL}. Fig.\ref{Fig3}a (b) displays the numerically time-dependent density distribution when the system is in topological (trivial) metal phase with $t_1=0.5 (1.2)$, $J=1$, $t_2=1$, and $L=60$ under open boundary condition. The corresponding time-averaged mean density displacements for the topological and trivial phases are shown in Fig.\ref{Fig3}c. Numeral calculations show that the time-averaged mean density displacement approaches the quantized value $0.5$ ($0.0$) for a large system with long $T$ using the finite-size analysis, which is attributed to the appearance of topological (trivial) states and indicates the system is in the topological (normal) metal phase. 

\section{2D topological metal}
Motivated and encouraged by the above-mentioned results, we straightway apply a similar analysis for the 2D case. Firstly the real-space Hamiltonian can be written as
\begin{eqnarray}
H&=\sum_{i,j}^{L_y,L_x}J(A_{i,j}^{\dagger}A_{i,j+1}+B_{i,j}^{\dagger}B_{i,j+1})+t_{1}A_{i,j}^{\dagger}B_{i,j}\nonumber \\ 
&+t_{1}A_{i,j}^{\dagger}B_{i-1,j}+t_{2}A_{i,j+1}^{\dagger}(B_{i,j}+B_{i-1,j})+H.c., 
\end{eqnarray}
where $A_{i , j}^{\dagger}$ ($B_{i , j}^{\dagger}$) and $A_{i , j}$ ($B_{i , j}$) are the fermionic creation and annihilation operators of the odd (even) quantum wire $i$ at lattice site $j$, respectively. The number of odd (even) quantum wires is $L_y$ and the length of the quantum chain is $L_x$. The primitive lattice vectors are defined as $a_1=(1,0)$ and $a_2=(0,1)$. For the reciprocal lattice, the corresponding primitive lattice vectors are $b_1=2\pi(1,0)$ and $b_2=2\pi(0,1)$. The Hamiltonian in momentum space can be written as
\begin{equation}
H=\left(
\begin{array}[c]{cc}
2J\cos k_x & (1+e^{ik_y}) (t_1+t_2e^{ik_x}) \\
(1+e^{-ik_y})(t_1+t_2e^{-ik_x})  & 2J\cos k_x
\end{array}
\right)\label{H2d}
\end{equation}

\begin{figure}[t]
\centering
\includegraphics[width=\columnwidth]{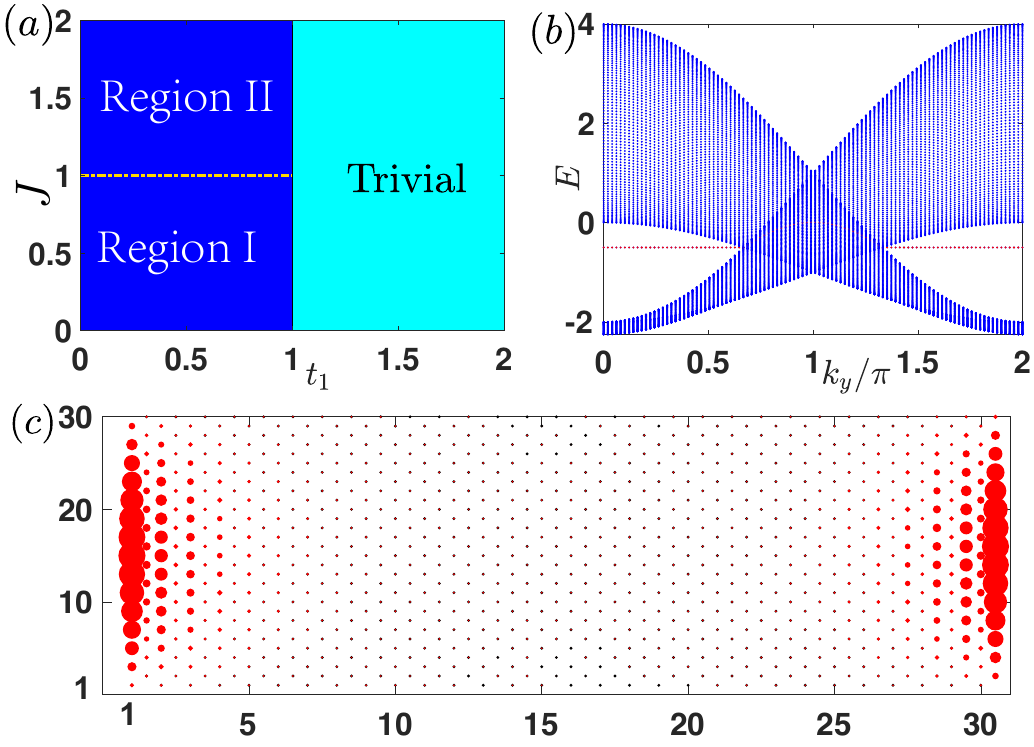}
\caption{(a) The phase diagram of 2D quantum wire array. Region I (II) represents that the system is topological metal with (without) edge states. (b) Energy spectra with system size $L_x=100$, $t_2=1$, and $t_1=J=0.5$ under open boundary condition in $x$ direction and periodic boundary condition in $y$ direction. The red dots in the energy band gap represent the topological nontrivial edge states. (c) The spatial distribution of a topological nontrivial edge state under open boundary condition in both directions. Other parameters are $L_x=2L_y=30$, $t_2=1$, and $t_1=J=0.5$.}
\label{Fig4}
\end{figure}

The Hamiltonian obeys the inversion symmetry $\mathcal{I}H(\mathbf{k})\mathcal{I}^{-1}=H(-\mathbf{k})$ similar to the three-leg ladder case, where inversion symmetry matrix $\mathcal{I}=\sigma_x$ in terms of Pauli matrices. According to the values of the topological invariant and behaviors of the topological edge states, the phase diagram can be described as shown in Fig.\ref{Fig4}a. In region I (II), the system is in topological metal phase with (without) the topological edge states. For the interchain couplings $t_1>t_2$, the system enters into the trivial normal metal phase. To show the topological edge states in region I, we plot the energy spectrum under open boundary condition in $x$ direction in Fig.\ref{Fig4}b. Clearly, the edge states (red dots) appear in the energy band gap for some momentum. The spatial distribution of a topological nontrivial edge state is shown Fig.\ref{Fig4}c.

Like the three-leg ladder case, the intrachain hopping $J$ modifies the energy spectra but not the corresponding eigenfunctions of the bulk Hamiltonian in Eq.\ref{H2d}. Thus, the topological property of topological metal is not changed and the energy gap closing could occur without bound touching because of the indirect gap. However, in a finite system, the amplitude of intrachain hopping could affect the spatial profile of the topological nontrivial edge states, which become delocalized for the strong intrachain hopping $J$. For the topological region II in Fig.\ref{Fig4}a, numerical calculations reveal that the time-averaged mean displacement in the $x$ direction approaches to the value 0.5, which indicates that the system enters the topological metal phase without edge states.

\section{Summary and outlook}
In short, we use a three-leg ladder and 2D lattice spinless quantum wire arrays sliding against each other as paradigmatic examples and demonstrate that alternated slide construction provides an elegant mechanism toward engineering topological metals. Remarkably, there is a topological metal without nontrivial edge states. The strategy of alternated slide construction could be experimentally confirmed in different systems such as photonic waveguide arrays, photonic crystals, topoelectrical circuits, or coupled acoustic resonators.

Finally, we would like to point out that the alternated slide construction can be a specific way to realize other topological phases of matters. For instance, the building-blocks could be extended to spin–orbit coupling (gap and gapless) systems, and topological chains (such as SSH model, Kitaev chain\cite{Kitaev01}, and AAH chain). Different types of topological insulators and superconductors can be constructed. On the other hand, the easily alternated slide construction also applies to multilayer systems including graphene moiré superlattices.

\emph{Acknowledgements.}-- We thank Weiyin Deng for helpful discussions. This work was supported by the National Natural Science Foundation of China (Grants No. 12074101, and 11604081). Z.W. Z. is also sponsored by the Natural Science Foundation of Henan (Grant No. 212300410040).


%

\appendix

\section{The energy dispersions of three-leg ladder}
\label{AppendixA}

\begin{figure}[hb]
\centering
\includegraphics[width=\columnwidth]{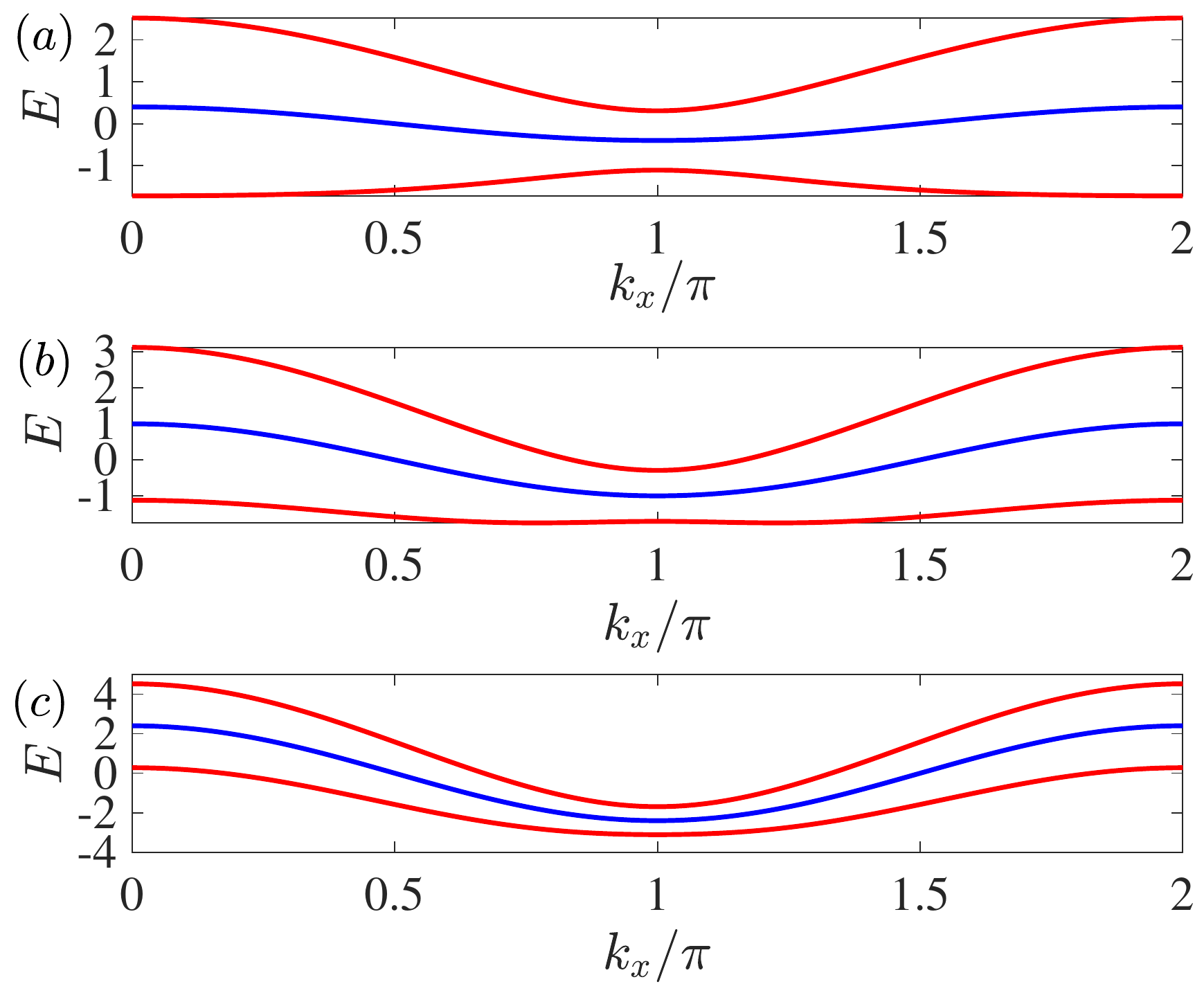}
\caption{Band structure of the three-leg ladder with intrachain couplings $t_1=0.5$, $t_2=1$ under periodic boundary condition. The blue line indicates the energy band of Hamiltonian $h_1$ and the red lines stand for the two energy band of Hamiltonian $h_2$. (a) Intrachain hopping $J=0.2$ case.  (b) $J=0.5$. (c) $J=1.2$.}
\label{Fig5}
\end{figure}

In this appendix, we investigate the energy structure for three-leg ladder in topological metal phases. Fig.\ref{Fig5} show the three energy dispersion curves $E_1$ and $E_{\pm}$ (Eqs. 5 and 6 in main text) for three different intrachain hoppings $J$ vs the Bloch wave number $k_x$ in momentum space when the interchain couplings $t_1=0.5$ and $t_2=1$. The blue line indicates the energy band $E_1$ of subspace Hamiltonian $h_1$ and the red lines stand for the two energy bands $E_{\pm}$ of subspace Hamiltonian $h_2$. When the intrachain hopping $J$ ($<\sqrt2/2$) is small, the energy gap of subsystem $h_2$ is open. The subsystem changes from a direct bandgap insulator to an indirect bandgap insulator without phase transition (see Fig.\ref{Fig5}a-b). The entire system is still topological metal with nontrivial edge states. As the intrachain hopping $J$ increases, the energy gap of Hamiltonian $h_2$ could be closed. However, the two energy bands of $h_2$ do not touch due to the indirect nature of the bandgap (see Fig.\ref{Fig5}c). As a consequence, in a finite system, the two degenerate edge states of Hamiltonian $h_2$ would assimilate into the bulk bands and become delocalize. Thus, the conventional bulk-boundary correspondence breaks down. The whole system enters into the topological metal phase without the nontrivial edge states.

\end{document}